\documentclass[10pt, conference]{IEEEtran} 
\IEEEoverridecommandlockouts  

\usepackage[oldenum,olditem]{paralist}
\usepackage{cite}
\usepackage{xcolor}
\usepackage{tikz}
\AtBeginEnvironment{quote}{\quotefont\small}

\usepackage{color, colortbl}
\usepackage{url}
\usepackage{pifont}
\usepackage{placeins}
\usepackage{graphicx}
\usepackage{comment}
\usepackage{csquotes}

\definecolor{picolor}{RGB}{51,102,0}
\newcommand{\mycheck}[1]{{\color{picolor}{\ding{51}}}}

\newcommand{\eg}{\textit{e.g., }}
\newcommand{\cf}{\textit{c.f., }}

\title{\LARGE \bf
Named Service Networking as a primer for the Metaverse
}

\author{Paulo Mendes$^{1}$ 
\thanks{$^{1}$Airbus, Willy-Messerschmitt-Strasse 1, 82024 Taufkirchen, Germany
        {\tt\small \{paulo.mendes\}@airbus.com}}%
}

\begin{document}

\maketitle
\thispagestyle{empty}
\pagestyle{empty}
    
\begin{abstract}
Ubiquitous extended reality environments such as the Metaverse will have a significant impact on the Internet, which will evolve to interconnect a large number of mixed reality spaces. Currently, Metaverse development is related to the creation of mixed reality environments, not tackling the required networking functionalities. This article analyzes suitable networking design choices to support the Metaverse, proposing a new service-centric networking approach capable of incorporating low-latency data fetching, distributed computing, and fusion of heterogeneous data types over the Cloud-to-Thing continuum.
\end{abstract}

\section{Introduction}
\label{sec.introduction}

The Metaverse will be based on interconnected ubiquitous \textit{eXtended Reality} (XR) spaces encompassing immersive and interactive virtual and cyber-physical systems, networked over an Internet powered to provide users ubiquitous cyber-virtual experiences. Several Metaverse applications, such as smart manufacturing, remote health, and intelligent transportation systems will require new networking frameworks to interconnect autonomic virtual and physical agents encompassing real-time data gathering and analysis.

In spite the general understanding about the internetworking requirements of the Metaverse, most of the discussion is centered on topics such as interoperability and ownership (\eg Non-Fungible Tokens). However, there is a lack of discussion around the networking functionalities needed to transport new types of data (\eg video, sound, haptic, holography) and to support the interaction between virtual applications and cyber-physical systems.

Without a complementary discussion about networking, we may end up with a misalignment between the application layer aspirations and the capabilities of the networking functionality. For instance, expectations about Web3.0, also known as Semantic Web \cite{2023towards:pronaya}, related to enhanced interactions are different from how data is exchanged at the networking layer.

From a networking perspective it is expected that the Metaverse will challenge traditional client-server web models and centralized cloud-based models, due to the need to support interoperability among a large number of XR spaces with low latency, which has to be kept lower than the human perceptible limit. To reduce latency, the Metaverse may rely on \textit{Multi-access Edge Computing} (MEC) \cite{2020JesusMEC} allowing cyber-physical devices to utilize nearby communication infrastructures and computing resources \cite{2020MEC-assisted:Du}, while powering the intertwining between 3D virtual realms and physical systems.

In this context, this article starts by discussing MEC limitations to support Metaverse in a Cloud-to-Thing continuum scenario, and the role of information-centric and service-centric networking to mitigate MEC challenges around resource discovery, dynamic function placement, and distributed data caching. A new framework intertwining in-network computing and \textit{Information Centric Networking} (ICN) along the Cloud-to-Thing continuum is proposed. This new framework, called \textit{Software Defined Named Service Networking} (SD-NSN), is proposed around a novel \textit{Named Service Networking} (NSN) data plane that is combined with a \textit{Software Defined} (SD) management plane defined based on a \textit{Software Defined Networking} (SDN) service deployment strategy. The notion of \textit{Named Service} proposed in this paper is a completely new approach. While the \textit{Name Service} used in the Internet and by cloud platforms is an offline system that binds names to IP addresses to support packet forwarding based on the topological locations (IP addresses), the proposed \textit{Named Service} based framework does not do any offline resolution of service names to IP addresses. On contrary, it provides an inline mechanism to forward packets based on service names.

The remainder of the article goes as follows: Section \ref{sec.fromedgetonsn} analyzes suitable networking approaches to support the Metaverse. After this initial analysis, Section \ref{sec.sdnsn} introduces the proposed SD-NSN networking framework and Section \ref{sec.namedservices} describes the new Named Service scheme used to manage computational services built based on serverless microservices. Section \ref{sec.operation} provides a detailed description of the SD-NSN operation in terms of service deployment and service execution. Finally, Section \ref{sec.metaverse} shows how the SD-NSN framework can be used to support Metaverse applications in the Cloud-to-Thing continuum, while section \ref{sec:summary} summarizes the contribution of the article, pointing also to future research directions.

\section{From Edge Computing to Named Service Networking}
\label{sec.fromedgetonsn}

Current Metaverse platforms rely on web protocols and cloud services suffering from performance limitations when interconnecting XR spaces. Some of the challenges pass by consistent throughput to handle high resolution XR applications and fast response times to computational requests. These challenges may be mitigated by bringing computing and storage resources towards the edge of the network \cite{2022ultra:Cai}. Moreover, since Metaverse applications aims to operate on a decentralized manner, based on distributed ledger technology and Web3.0, it is importance that rely on a networking framework able to support decentralized services such as name resolution, identity management and resource discovery.

This section analyzes three networking frameworks capable of supporting low latency, while protecting user privacy in a decentralized network: Edge computing, information-centric networking and service-centric networking.

\subsection{Edge Computing}
\label{sec.fromedgetonsn.edgecomp}

Edge computing refers to enabling technologies, such as SDN, \textit{Network Function Virtualization} (NFV) and \textit{Service Function Chaining} (SFC), capable of supporting the execution of services at network edges. Processing services at the edge, and not at the cloud, brings advantages, such as overcoming the fluctuations of end-to-end bandwidth and delay, as well as protecting privacy, since data does not need to be uploaded to data centers.

Current edge computing frameworks focus on the edge-cloud relationship to achieve low delays and high throughput. Such edge computing frameworks do not look at scenarios in which different edges may cooperate between themselves, as well as with devices located in the far edge. In such a Cloud-to-Thing continuum, services may be computed also on mobile devices such as vehicles and satellites. 

The extension of the edge-cloud relationship towards a Cloud-to-Thing continuum aims at improving response times and energy savings \cite{2022elastic:sergio} to fulfill the requirements of Metaverse applications, while solving some of the limitations of edge computing, namely: (i) dependency on cloud systems when lacking resources at the edge, (ii) intermittently available edge services, due to unstable connectivity between mobile devices and edges, (iii) idle edge services when data is not exported to the edge due to privacy issues.

Moreover, the ultimate goal of the Metaverse is interoperability, allowing users to roam with their identities, assets, and data between XR spaces. Hence, there is the need to consider security at service and data levels, a functionality that may be provided by networking frameworks based on ICN. Moreover, ICN approaches can also support decentralized networking over the Cloud-to-Thing continuum while keeping edge computing properties such as caching.

\subsection{Information Centric Networking}
\label{sec.fromedgetonsn.ndn}

From a networking point of view, the Metaverse requires communication between data objects and services and not between devices, since the location of assets can change and Metaverse information may be associated with moving entities. In this scenario, a networking framework based on host identifiers and end-to-end host connectivity should be complemented with a data-based and service-based networking approach.

Hence, this article analyzes the usage of \textit{Named Data Networking} (NDN), a distributed manifestation of ICN \cite{2022Kua:varun}, to address the networking challenges raised by the Metaverse \cite{2022JeffStatement}, namely data persistence across distributed XR spaces and interoperability between nomadic Metaverse objects \cite{2021ZhangNDNMPS}.

At an architectural level, NDN can be seen as shifting HTTP request and response semantics to the network layer. With NDN, requests for named data operate at packet granularity; NDN forwards Interest packets based on named data, and forwards Data packets via the reverse path of Interest packets. The packet granularity creates a fine tuned feedback loop (\eg congestion, failures) useful for the adaptation of dynamic services and networks, as is the case of the Metaverse. Moreover, builtin communication security and authentication is provided by binding name and content in each data packet through a cryptographic signature.

Besides the builtin security and the capability to adapt to dynamic services and networks, NDN is also able to support other Metaverse requirements, such as group communication and synchronization, as well as session-oriented communication based on an ICN-aware RESTful protocol \cite{2022RestfulICN} that may leverage name privacy, which is essential for Metaverse applications.

However, while NDN is focused on data retrieval, it is expected that Metaverse applications will gain from a network able to support the execution of services while leveraging storage, networking and computing resources located along the Cloud-to-Thing continuum. This may be achieved based on a service-centric networking approach, in which case NDN data retrieval is just one possible service.

\subsection{Service Centric Networking}
\label{sec.fromedgetonsn.nsn}

There has been extensive discussion in the literature about the benefits of joining NDN and edge computing \cite{2020platforms:ioini} to bring computing to the network edge by leveraging resources of partially or entirely idle devices. These efforts are focused on the deployment of monolithic services that may require resources that are not available in devices deployed in the edge and far edge segments of the Cloud-to-Thing continuum.
 
To support Metaverse low latency requirements taking into account the constrained resource of heterogeneous devices in the Cloud-to-Thing continuum, a service-centric networking framework should be based on microservices executed as serverless functions inside selected devices. The motivation to look at serverless functions is related to their capability to simplify service management on heterogeneous devices \cite{2020platforms:ioini}.

This means that NDN needs to be leveraged to support also the storage and execution of correlated serverless microservices. 

NDN has been proven to be a good basis for the development of in-network computing solutions. NFN \cite{2016NFN:Scherb}, NFaaS \cite{2017nfaas:krol} and CFN \cite{2019CFN:krol} are examples of in-network computing frameworks based on NDN. The concepts of NFN and NFaaS support the execution of stateless functions, while CFN supports the execution of stateful functions. The approach followed by CFN may bring limitations to manage data replication across the network, namely to ensure adaptation during run-time. Despite their stateless or stateful approach, neither NFN, NFaaS nor CFN support the execution of services requiring more elaborated processing, such as chains of computational functions. 

Going beyond prior art, this article argues that Metaverse can be supported in the Cloud-to-Thing continuum by a novel \textit{Software-Defined Named Service Networking} (SD-NSN) framework, in which a \textit{Software Defined} (SD) management plane able to deploy correlated serverless microservices is combined with a new \textit{Named Service Networking} (NSN) data plane able to route packets based on names describing services composed of chains of computational functions. The proposed SD-NSN framework is capable of executing services defined based on serverless microservices in a decentralized manner.

\section{Software Defined Named Service Networking Framework}
\label{sec.sdnsn}

Managing services deployed as a set of serverless microservices in the Cloud-to-Thing continuum requires the ability to continuously update their placement in the network. Hence, a \textit{Software Defined} (SD) management plane is used to deploy chains of serverless microservices on demand and to automatically relocate them reacting to network changes, while respecting their interdependencies in order to support the execution of the respective service.

\begin{figure}[hbt!]
    \centering
    \includegraphics[width=.5\textwidth]{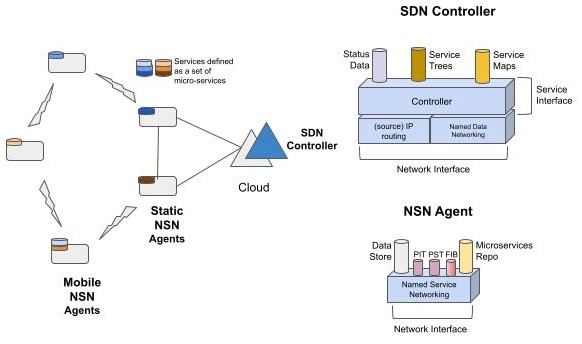}
    \vspace{-1em}
    \caption{Software Define Named Service Networking framework}
    \label{fig:buildingblocks}
\end{figure}

The SD management plane is implemented by an SDN controller installed in a data center or on an edge device, such as a satellite ground station, following a common MEC practice in which computing resources are made available at the edge of a network, by using mini-data center solutions such as the Microsoft Azure Modular Datacenter. The SDN controller manages the deployment of serverless microservices in NSN agents installed in any device of the Cloud-to-Thing continuum, as illustrated in Figure \ref{fig:buildingblocks}. For that, it gathers services developed by third-parties and decides about in which network devices to install each of the serverless microservices of the provided service. This decision is supported by information periodically collected from the network. Such information is also used to decide about relocating serverless microservices inside the network. The SD management plane is described in section \ref{sec.sdnsn.controlplane}.

NSN agents implement the proposed new NSN protocol stack, created based on the NDN data plane with the addition of extra data structures and a new forwarding strategy, as described in section \ref{sec.sdnsn.dataplane}. Besides the novel NSN data plane, NSN agents also implement a local proxy to translate service requests received from local consumers. These requests can have the format of a \textit{Service Request} as defined in the \textit{Routing on Service Addresses} proposal \cite{2023rosa:trossen}, in which case the NSN agent operates as a \textit{Service Instance}.

\subsection{SD Management plane}
\label{sec.sdnsn.controlplane}

The SDN controller makes use of two interfaces: a network interface to interact with devices and a service interface to interact with developers. To support its operation, the SDN controller uses two new data structures - named service charts and service trees - to handle Named Services as defined in section \ref{sec.namedservices}.

As illustrated in Figure \ref{fig:buildingblocks}, the network interface encompasses an NDN and an TCP/IP protocol stack. The NDN protocol stack is used to gather network information as well as service requests from NSN agents. The TCP/IP protocol stack is used to install serverless microservices in different devices. The usage of an NDN protocol stack allows the SD management plane to make full use of NDN in-network caching to obtain network status from any intermediate node and not only from the network node that originated that status data, helping to increase reliability and decrease latency. Since in-network caching and innate multicast capability of NDN decreases the overall data transmission, a higher network throughput may be achieved.

Through the NDN stack, the SDN controller uses the following services: (i) N-to-1, allowing it to collect status information from network nodes, being this service defined by a name prefix \textit{/sdn-nsn/monitor}, and (ii) 1-to-1, allowing NSN agents to gather information from the SDN controller about the chains of serverless microservices needed to execute specific services, being this service identified by the name prefix \textit{/sd-nsn/retrieve/servicehead}.

Using the NDN stack to deploy serverless microservices would lead to extra overhead, due to the push nature of NDN and the pull nature of the deployment operation. Instead, the SDN controller implements a novel publish primitive to install microservices in NSN agents based on an IP source based forwarding approach that uses the IP addresses collected while monitoring the network. The source based forwarding approach is implemented based on a new packet type able to carry serverless microservices in the payload and source routing information in the packet header. The routing information is generated by the SDN controller based on the network topology gathered during the monitoring operation.

Due to the dynamic nature of the Metaverse and of the Cloud-to-Thing continuum, the association of serverless microservices to specific NSN agents is done by a decision making process that uses information about the services to be deployed and information about the network status. In what concerns services, the decision making process considers their execution time, the execution time of each of their microservices, and the correlation between microservices. In what concerns network status, decisions are made taken into account the delay between each NSN agent, the storage, computing and energy capacity of each agent, as well as the distance between agents and data sources.

A detailed explanation about the operation of the SD management plane is provided in section \ref{sec.operation.deployment}.

\subsection{NSN Data Plane}
\label{sec.sdnsn.dataplane}

The proposed NSN data plane extends the NDN architecture by integrating: (i) a novel named service scheme, described in section \ref{sec.namedservices}, (ii) a new data structure to track pending service requests, (iii) a service store to host serverless microservices, and (iv) a named service forwarding engine to process all interests and data packets that have the name prefix \textit{/sd-nsn}, in analogy to the way NDN handles name checksum hashes.

In terms of data structures, besides the \textit{Pending Interest Table} (PIT) already present in the NDN node architecture, NSN uses a new \textit{Pending Service Table} (PST) to register the names of the services that were received in Interest packets and that were not serviced yet. While the PIT is used to point a data name to an interface over which Data packets should be forwarded, a PST entry is used to point a service name to a local microservice. Since the SD management plane follows a source-routing approach, the NSN data plane does not have a \textit{Forwarding Information Base} (FIB) as exists in NDN. As shown in section \ref{sec.namedservices}, this means that packets are forwarded based on service names encompassing information about the chain of microservices, each of which potentially installed in a different device, that need to be execute in an order defined by the service.

NSN stores serverless microservices in a local code repository, which includes a queue of all microservices that are being executed. The registration of code functions is done based on the SD management plane publish primitive. Since NSN follows an SDN approach, contrary to what happens with NFN, the installation of code functions does not require populating a FIB with a corresponding namespace entry.

A detailed explanation of the operation of the NSN data plane is provided in section \ref{sec.operation.execution}.

\section{Named Services}
\label{sec.namedservices}

\begin{figure}[hbt!]
    \centering
    \includegraphics[width=.5\textwidth]{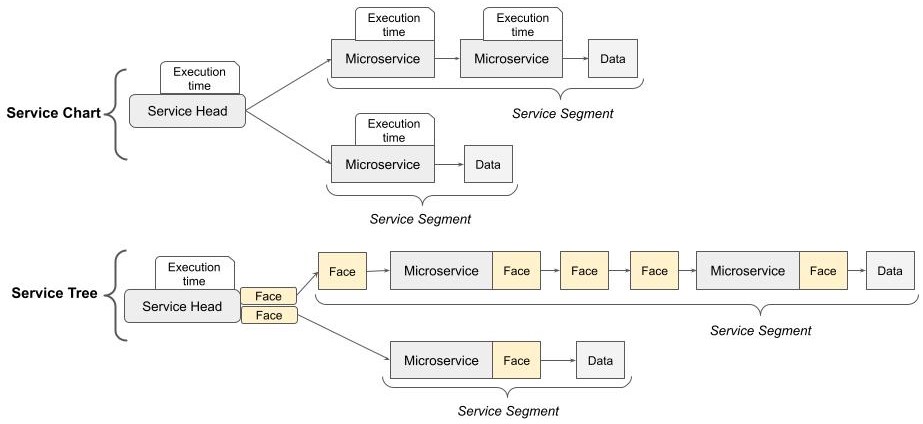}
    \vspace{-1em}
    \caption{Service Chart and Service Tree}
    \label{fig:servicemapssegments}
\end{figure}

Services are defined based on microservices that are intertwined in a specific order to reflect their execution priorities. Hence, developers should define services based on Service Charts describing the relationship between microservices, as shown in Figure \ref{fig:servicemapssegments}.

A Service Chart may encompass different segments to allow the description of services that require the fusion of different data types, such as video and audio. In a Service Chart, each segment encompasses two blocks: microservices and data.

Service developers register Service Charts near the SDN controller, while customers request services near the closest NSN Agent by referencing only to Service Heads.

The SDN controller uses Service Charts and information retrieved from the netwok to define Service Trees, which encompass the information needed for the execution of services based on correlated serverless microservices. A Service Tree encompasses several service segments (\cf figure \ref{fig:servicemapssegments}) with the code of microservices, and an indication about the interfaces (face in NDN terminology) to be used by an NSN Agent to reach the NSN Agent where the next microservice is deployed. The decision about the interfaces to use is done based on the topological data retrieved by the SDN controller, such as node neighbourhood and link costs.

Since the NSN data plane is based on NDN, Service Trees are described based on hierarchical names. For instance, a multimedia service merging video and sound may be represented by a Service Tree with two service segments, each one chaining several microservices, as well as other segment levels. The following example illustrates a multimedia service with one segment level, S1, encompassing two segments, S11 for video and S12 for audio, each of which includes two microservices: data gathering (/videoaircraft320 and /soundfactory) and data analysis (/videoanalysis and /soundanalysis):

\begin{displayquote}
        \textit{/multimedia/S11/face1/videoanalysis/face30/videoaircraft320/
    S12/face2/soundanalysis/face10/soundfactory}
\end{displayquote}

The proposed named service scheme identifies not only microservices, but also the interfaces that need to be crossed to transit between microservices. This source routing approach aims to reduce complexity by avoiding the implementation of a name-based routing protocol, as well as FIBs.

In a Service Tree name, \textit{faceX} refers to an interface that needs to be crossed to reach the next microservice or data source. For instance, in the above example, \textit{face30} refers to the interface that the NSN agent running the /videoanalysis microservice needs to use to forward Interest packets to the NSN agent that implements the microservice /videoaircraft320.

\section{SD-NSN Operation}
\label{sec.operation}

\begin{figure}[hbt!]
    \centering
    \includegraphics[width=.5\textwidth]{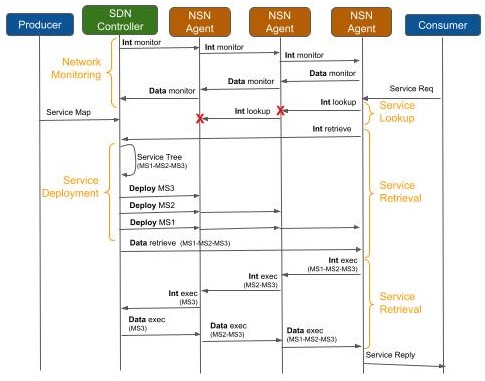}
    \vspace{-1em}
    \caption{SD-NDN Signalling}
    \label{fig:SD-NDN Signalling}
\end{figure}

The SD-NDN operation encompasses two parallel phases, as illustrated in Figure \ref{fig:SD-NDN Signalling}: service deployment and service execution. The former is triggered when the SDN controller receives charts of services to be deployed, and includes two operations as described in section \ref{sec.operation.deployment}: network monitoring and service placement. The service execution phase is triggered when a service request is received by an NSN agent via the local proxy including the operations described in section \ref{sec.operation.execution}; service lookup, service retrieval and service forwarding.

\subsection{Service Deployment}
\label{sec.operation.deployment}

\subsubsection{Network Monitoring}
\label{sec.operation.deployment.mon}

The SDN controller monitors the network by sending periodic Interest packets with name prefix \textit{/sd-nsn/monitor}. NSN agents do not parse this name prefix using the NSN forwarding engine but using the NDN engine. When the SDN controller is deployed in the cloud, the Interest packets are encapsulated within IP packets having the address of the edge of the NSN network as destination address.

When an NSN agent gets an Interest packet it performs the following actions: (i) If there are more interfaces leading to other NSN agents, it forwards the Interest packet over all those interfaces, (ii) otherwise the NSN agent terminates the Interest packet and creates a Data packet with the requested status data such as: battery lifetime, available storage, average delay to neighbor devices. This information is stored in the content store with prefix \textit{/sd-nsn/monitor}.

When an NSN agent gets a Data packet with name \textit{/sd-nsn/monitor} it adds to the packet the local status information, after which it forwards the packet following the information stored in the PIT.

As a result of the monitoring phase, the SDN controller creates an image of the network topology.

\subsubsection{Service Placement}
\label{sec.operation.deployment.deploy}

\begin{figure}[hbt!]
    \centering
    \includegraphics[width=.5\textwidth]{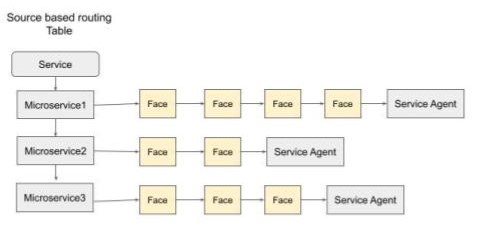}
    \vspace{-1em}
    \caption{Source based routing table}
    \label{fig:sourceroutingtable}
\end{figure}

The SDN controller uses the network image resulting from the monitoring phase and the Service Charts retrieved through the developer interface to decide in which NSN agents to install microservices. By combining this information the SDN controller creates entries in a source routing table, illustrated in Figure \ref{fig:sourceroutingtable}, used by the publish primitive to install microservices in different NSN agents based on an IP source based forwarding approach.

The publish primitive encompasses a new packet type, called \textit{Deploy}, which carries the microservice to be installed in the payload, while the packet header carries information about all interfaces that need to be crossed to reach the NSN agent where the microservice needs to be installed.

When an NSN agent receives a \textit{Deploy} packet, it reads the first element of the chain carried in the packet header, and places the \textit{Deploy} packet in the outgoing queue of that interface, after removing it from the packet header. This process continues until the chain in the \textit{Deploy} packet header is reduced to one element, meaning that the microservice should be installed in the current NSN agent.

\subsection{Service Execution}
\label{sec.operation.execution}

\subsubsection{Service Lookup}
\label{sec.operation.execution.lookup}

This process is triggered in an NSN agent when an user requests the execution of a service via an IP/ICN proxy. After receiving a service request, the NSN agent checks if the Service Head of the requested service is found locally. If so, the service execution is triggered; otherwise, the NSN agent starts a service lookup mechanism based on an Interest packet with name:

\begin{displayquote}
 \textit{/sd-nsn/lookup/servicehead/executiontime/triggeringtime}
\end{displayquote}

When an NSN agent receives such an Interest packet, it performs a lookup in the local microservice repository. If the service head is found, the service execution is triggered, if and only if the service execution time plus the Interest packet delay, measured as the difference between the current locally measured time and the \textit{triggeringtime} indicated in the Interest name, is lower than the \textit{executiontime} carried in the Interest name. 

If the service execution is triggered, a Data packet with the service result is forwarded back through the inverse path of the Interest packet, as provided by the PIT. However, if the service head is not found locally, the Interest packet is forwarded to all neighbors following the NDN multicast forwarding strategy. If the Interest packet cannot be forwarded, it is dropped and no Data packet is generated.

If the NSN agent that triggered the Service Lookup mechanism does not receive any Data packet within a stipulated timeout (\eg equal to two times the service execution time), it concludes that the requested service is not deployed and it triggers the Service Retrieval method.

\subsubsection{Service Retrieval}
\label{sec.operation.execution.retrieval}

In order to retrieve a Service Tree, an NSN agent needs to create an Interest packet with name prefix \textit{/sd-nsn/retrieve/servicehead}. The Interest packet is forwarded based on the NDN FIB that might already have a pre-configured path to the SDN controller, or based on the default NDN multicast forwarding strategy.

When the SDN controller receives such an Interest packet, it triggers the Service Definition method, after which the resulting Service Tree is sent back in a Data packet that follows the inverse path of the correspondent Interest packet. Afterwards, the SDN controller triggers the Service Deployment method based on the created Service Tree. When the NSN agent that sent the service retrieval Interest packet gets the Data packet, it uses the name of the carried Service Tree to trigger the Service Forwarding method.

\subsubsection{Service Forwarding}
\label{sec.operation.execution.execution}

The execution of a service requires running all microservices defined in its Service Tree, by coordinating a chain of Interest/Data packets forwarded between NSN agents implementing the different microservices. The execution of a chain of microservices is done following a source routing approach in which the Service Tree is integrated in Interest packets, with name prefix \textit{/sd-nsn/exec}, which are forwarded based on the information carried in the service name.

Since the size of Interest packets depends upon the used names, transmission overheads can be mitigated by reducing the size of service names. This is achieved by updating the name carried in Interest packets when they cross NSN agents. For instance, the reception of an Interest packet with name \textit{/videoanalysis/face30/video-aircraft320} by an NSN agent implementing the microservice \textit{/videoanalysis} leads to the transmission of an Interest packet with name \textit{/video-aircraft320}, being the prefix \textit{/videoanalysis} stored in the local PST, while the prefix \textit{/face30} corresponds to the face to be used to forward the new Interest packet. 

\begin{figure}[hbt!]
    \centering
    \includegraphics[width=.5\textwidth]{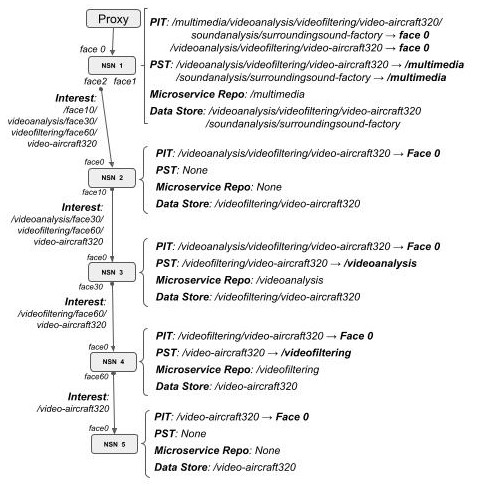}
    \vspace{-1em}
    \caption{Service execution example}
    \label{fig:serviceexecution}
\end{figure}

To better understand the operation of the NSN data plane, Figure \ref{fig:serviceexecution} provides an example of the execution of the following multimedia service over a set of five different devices:

\begin{displayquote}
    \textit{/sdn-ndn/exec/multimedia/S11/face1/videoanalysis/face30/
    videoaircraft320/S12/face2/soundanalysis/face10/soundfactory}
\end{displayquote}

In general terms, an NSN agent perform the following operations upon the reception of an Interest packet with name prefix \textit{/sd-nsn/exec} (the operation of node NSN3 in Figure \ref{fig:serviceexecution} is used as example):
\begin{enumerate}
   \item Verifies if the requested microservice (\textit{/videoanalysis}) is in the Microservice repository;
   \item Includes in the PIT the face (face 0) over which the Interest was received, if the microservice was found in the repository;
   \item Checks if the data requested by the microservice is available in the data store (\eg \textit{/videofiltering/video-aircraft320});
   \item If it is, executes the microservice and sends back the result in Data packets via all faces indicated in the PIT;
   \item If it is not:
   \begin{enumerate}
         \item Identifies the service segments that include microservices to be executed;
         \item Creates a new PST entry with the identified name segment (\eg \textit{/videofiltering/video-aircraft320}) pointing to locally stored microservices (\textit{/videoanalysis}) that depend upon the execution of the microservices in that segment;
         \item Terminates the Interest packet and creates other Interest packets with the name of the microservices stored in the PST;
         \item Sends the new Interest packets via the faces indicated in the service segments (\eg face 30), carrying a name that is equal to the one carried by the terminated Interest packet, after removing the name of the local microservices and local faces (\eg \textit{/videofiltering/face60/video-aircraft320}).
   \end{enumerate}
\end{enumerate}

As occurs in NDN, when an Interest packet reaches an NSN agent that has the needed data (locally gathered or resulting from the execution of a local microservice), a Data packet is created with the same name prefix as the one received in the Interest packet. The Data packet is sent via all faces indicated in the PIT. When an NSN agent receives a Data packet, it performs the following operations:
\begin{enumerate}
  \item Stores the received data in the Data store;
  \item Verifies if some neighbors need the received data by checking the PIT, in which case the Data packet is forwarded via the faces indicated in the PIT;
  \item Verifies if some local microservices need the received data, by checking the PST, in which case the data carried in the packet is passed to the local microservices, and the Data packet is terminated.
  \item If a local microservice is executed, the result is included in a new Data packet with the name of the local executed microservice plus the name of all used data objects: for instance \textit{/videoanalysis/video-aircraft320} as a result of the execution of the \textit{/videoanalysis} microservice based on the data \textit{/video-aircraft320} carried in the received Data packet. The new Data packet is forwarded via all faces indicated in the relevant PIT entry.
\end{enumerate}

\section{Support of Metaverse Applications}
\label{sec.metaverse}

As we move toward a Cloud-to-Thing environment, the concept of SD-NSN is a promising service-centric networking approach supporting the Metaverse. It allows dynamic networks to create and adapt service segments in run-time, while supporting the basic functionalities of the Metaverse: decentralization and service fusion. Metaverse applications can take advantage of the fusion of different services such as \textit{Artificial Intelligence} (AI) and Blockchain \cite{2022Fusion:Yang} while exploiting data from different sources to regenerate a comprehensive view of the virtual-physical environment.

For example, in an industrial digital twin scenario, to warn a robot about a hazardous situation in an hidden location of a factory, NSN agents deployed in the factory can compute an alert service by fusing a video processing service based on images collected by different cameras deployed close to the hidden zone, with a sound analysis process based on audio samples collected from a set of microphones deployed close by.

\section{Summary}
\label{sec:summary}

This article proposes the \textit{Software Defined Named Service Networking} (SD-NSN) framework, leveraging the information-centric networking paradigm to create a service-centric networking approach that allows the deployment and execution, in a Cloud-to-Thing continuum, of services defined based on serverless microservices. The proposed SD-NSN framework brings the flexibility needed to support the basic functionalities of the Metaverse, decentralization and service fusion, based on a novel \textit{Named Service Networking} (NSN) data plane, which uses a new Named Service scheme to describe services as a chain of serverless microservices deployed based on a \textit{Software Defined Networking} (SDN) strategy. The article describes the operation of the SD-NDN framework encompassing two parallel phases: service deployment and service execution. Future work encompasses the evaluation of the SD-NSN framework in a Cloud-to-Thing continuum scenario including the implementation of NSN agents in satellites and ground stations, while the SDN controller can be implemented in a cloud center, or based on a distributed system of edge devices (\eg satellite ground stations).

\bibliographystyle{IEEEtran}
\bibliography{metanet}

\begin{thebibliography}{10}
\providecommand{\url}[1]{#1}
\csname url@samestyle\endcsname
\providecommand{\newblock}{\relax}
\providecommand{\bibinfo}[2]{#2}
\providecommand{\BIBentrySTDinterwordspacing}{\spaceskip=0pt\relax}
\providecommand{\BIBentryALTinterwordstretchfactor}{4}
\providecommand{\BIBentryALTinterwordspacing}{\spaceskip=\fontdimen2\font plus
\BIBentryALTinterwordstretchfactor\fontdimen3\font minus
  \fontdimen4\font\relax}
\providecommand{\BIBforeignlanguage}[2]{{%
\expandafter\ifx\csname l@#1\endcsname\relax
\typeout{** WARNING: IEEEtran.bst: No hyphenation pattern has been}%
\typeout{** loaded for the language `#1'. Using the pattern for}%
\typeout{** the default language instead.}%
\else
\language=\csname l@#1\endcsname
\fi
#2}}
\providecommand{\BIBdecl}{\relax}
\BIBdecl

\bibitem{2023towards:pronaya}
P.~Bhattacharya, D.~Saraswat, D.~Savaliya, S.~Sanghavi, A.~Verma, V.~Sakariya,
  S.~Tanwar, R.~Sharma, M.~S. Raboaca, and D.~L. Manea, ``Towards future
  internet: The metaverse perspective for diverse industrial applications,''
  \emph{MDPI Mathematics}, Feb. 2023.

\bibitem{2020JesusMEC}
J.~M. et~al., ``Harmonizing standards for edge computing - a synergized
  architecture leveraging etsi isg mec and 3gpp specifications,'' \emph{ETSI
  White Paper 36}, Jul. 2020.

\bibitem{2020MEC-assisted:Du}
J.~Du, F.~R. Yu, G.~Lu, and J.~Wang, ``Mec-assisted immersive vr video
  streaming over terahertz wireless networks: A deep reinforcement learning
  approach,'' \emph{IEEE Internet of Things Journal}, Jun. 2020.

\bibitem{2022ultra:Cai}
Y.~Cai, J.~Llorca, A.~M. Tulino, and A.~F. Molisch, ``Ultra-reliable
  distributed cloud network control with end-to-end latency constraints,''
  \emph{arXiv:2205.02427}, May 2022.

\bibitem{2022elastic:sergio}
S.~L. et~al, ``Elastic data analytics for the cloud-to-things continuum,''
  \emph{IEEE Internet Computing}, Dec. 2022.

\bibitem{2022Kua:varun}
V.~Patil, H.~Desai, and L.~Zhang, ``Kua: A distributed object store over named
  data networking,'' in \emph{ACM Conference on Information-Centric
  Networking}, Osaka, Japan, September 2022.

\bibitem{2022JeffStatement}
J.~Burke, ``Statement: As tcp/ip is to the web, icn is to the...?'' in
  \emph{ACM Conference on Information-Centric Networking}, Osaka, Japan,
  September 2022.

\bibitem{2021ZhangNDNMPS}
Z.~Zhang, S.~Liu, R.~King, and L.~Zhang, ``Ndn-mps: Supporting multiparty
  authentication over named data networking,'' in \emph{ACM Conference on
  Information-Centric Networking}, Paris, France, September 2021.

\bibitem{2022RestfulICN}
D.~Kutscher and D.~Oran, ``Restful information-centric networking: statement,''
  in \emph{ACM Conference on Information-Centric Networking}, Osaka, Japan,
  September 2022.

\bibitem{2020platforms:ioini}
N.~E. Ioini, D.~H{\"a}stbacka, C.~Pahl, and D.~Taibi, ``Platforms for
  serverless at the edge: a review,'' in \emph{Springer European Conference on
  Service-Oriented and Cloud Computing}, Wittenberg, Germany, Mar. 2020.

\bibitem{2016NFN:Scherb}
C.~Scherb, M.~Sifalakis, and C.~Tschudin, ``A packet rewriting core for
  information centric networking,'' in \emph{IEEE Consumer Communications and
  Networking Conference}, Las Vegas, USA, January 2016.

\bibitem{2017nfaas:krol}
M.~Kr{\'o}l and I.~Psaras, ``Nfaas: named function as a service,'' in \emph{ACM
  Conference on Information-Centric Networking}, Berlin, Germany, September
  2017.

\bibitem{2019CFN:krol}
M.~Krol, S.~Mastorakis, D.~Oran, and D.~Kutscher, ``Compute first networking:
  Distributed computing meets icn,'' in \emph{ACM Conference on
  Information-Centric Networking}, Macau, China, September 2019.

\bibitem{2023rosa:trossen}
D.~Trossen, L.~Contreras, J.~Finkhaeuser, and P.~Mendes, ``Routing on service
  addresses,'' in \emph{IETF draft draft-trossen-rtgwg-rosa-02}, February 2023.

\bibitem{2022Fusion:Yang}
Q.~Yang, Y.~Zhao, H.~Huang, Z.~Xiong, J.~Kang, and Z.~Zheng, ``Fusing
  blockchain and ai with metaverse: A survey,'' \emph{IEEE Open Journal of the
  Computer Society}, Jan. 2022.

\end{thebibliography}

\section*{Biographies}

\textbf{Paulo Mendes} (PhD 2004) is Expert in Communication Network Architectures and Design at Airbus (Munich, Germany). Paulo holds a Ph.D. in Informatics Engineering from University of Coimbra, having done his research work as invited researcher at Columbia University. He is also Associate Researcher at the Technical University of Munich and at the iSTAR-IUL research center. Paulo holds over 80 publications, 10 book chapters and 18 patents. He is an ACM member and an IEEE senior member.

\end{document}